\documentclass[aps,prl,preprint,groupedaddress]{revtex4}

\usepackage{graphicx}
\usepackage{subfigure}
\usepackage{bm}
\usepackage{ams}

\begin{document}


\title{SPARTAN RANDOM PROCESSES IN TIME SERIES MODELING}


\author{M.~\v{Z}ukovi\v{c}}
\email[]{mzukovic@mred.tuc.gr}
\affiliation{Technical University of Crete}

\author{D.~T. Hristopulos}
\email[]{dionisi@mred.tuc.gr}
\affiliation{Technical University of Crete}


\date{\today}

\begin{abstract}
A Spartan random process (SRP) is used to estimate the correlation
structure of time series and to predict (extrapolate) the data
values. SRP's are motivated from statistical physics, and they can
be viewed as  Ginzburg-Landau models. The temporal correlations of
the SRP are modeled in terms of `interactions' between the field
values. Model parameter inference employs the computationally fast
modified method of moments, which is based on matching sample
energy moments with the respective stochastic constraints.  The
parameters thus inferred are then compared with those obtained by
means of the maximum likelihood method. The performance of the
Spartan predictor (SP) is investigated using real time series of the
quarterly S\&P 500 index. SP prediction errors are compared with
those of the Kolmogorov-Wiener predictor. Two predictors, one of
which explicit, are derived and used for extrapolation.  The
performance of the predictors is similarly evaluated.
\end{abstract}


\maketitle

\section{Introduction}

Time series that carry information about temporal autocorrelations
in variables such as stock prices, interest rates, etc. have wide
applications in finance \cite{chan}. Such information allows
predictions of time-series, estimating the associated prediction
uncertainty, and performing stochastic simulations for
reconstructing realizations of the process. For Gaussian time
series, the temporal structure is estimated from the data by
calculating the autocovariance matrix or the structure function
(variogram). A different approach, typically used in statistical
physics, focuses on physical interactions embodied in the energy
functional.

Statistical physics plays an increasingly important role in
economics, helping to understand the behavior of complex economical
and financial systems (\cite{econ_ts}). However, applications in the
prediction and simulation of such systems are explored less.
Recently, a class of Gaussian
 random fields, named Spartan Spatial Random Fields
(SSRF) \cite{dth03} was proposed as a general framework for spatial
modelling. Its main advantages lie in parametric frugality,
potential for including physical constraints in the probability
density function (pdf), and efficient model estimation. Herein we
define Spartan Random Processes (SRP) in time, formulate a Spartan
predictor (SP), and investigate its potential for time series
prediction.

\section{Spartan Random Processes}
\label{spartan}

Let us consider a noise-free detrended time-series
${X_{\lambda}}({t})$ that represents the fluctuations of an
observable with temporal resolution $\lambda$. For statistical
inference from a single realization (state), second-order
stationarity \cite{yaglom} and ergodicity \cite{adler,lantu} are
often assumed. The latter implies that the `characteristic' scale of
the fluctuations be considerably smaller than the domain size. In
statistical physics, the pdf of {\it stationary Gaussian time
series} can be expressed in terms of an energy functional $
{H[X_\lambda({t})]} $, according to the Gibbs pdf $ f_{\rm x}
[X_\lambda] = Z^{- 1} \exp \left\{ { - H[X_\lambda]} \right\},$
where $ Z $ is a normalizing constant (the partition function).

The SRPs can be defined following the formalism introduced in
\cite{dth03}, restricted to one dimensional domains. The
fluctuation-gradient-curvature (FGC) SSRF model introduced in
\cite{dth03} embodies Gaussian fluctuations and involves three terms
that measure the square of the magnitude, the gradient and the
curvature of the fluctuations. On a 1D chain, the FGC form of
$H[X_\lambda]$ can generally be written as

\begin{equation}
\label{fgc_1D} H_{\rm fgc} [X_\lambda ; \bm{\theta} ] = \frac{1}{{2\eta _0 \xi
}}\int {dt} \, \left\{ \left[ {X_\lambda (t)} \right]^2
+ \eta _1 \,\xi ^2 [ {\dot{X}_\lambda (t)} ]^2  + \xi ^4 [
{\ddot{X}_\lambda (t)}]^2 \right\},
\end{equation}

\noindent where $\dot{X}_\lambda (t)$ and $ \ddot{X}_\lambda (t)$
denote respectively the first and second time derivatives, and
$\bm{\theta}=(\eta_0,\eta_1,\xi,k_c)$ is a vector of model
parameters: $\eta_0$ is the scale coefficient,  $\eta_1$ the
autocovariance shape coefficient,  $\xi$ is the characteristic
length, and  $k_c \propto \lambda ^{-1}$ the cutoff frequency. For
discrete time series sampled at $t_{i}=i \alpha$, $i=1,\ldots,N$,
$\alpha
>0$, the energy functional can be expressed in terms of local
energies $S_{m}(t_{n}),\ m=0,1,2$, as follows:
\begin{equation}
\label{H1dS} H_{{\rm{fgc}}} \left[ X_{\lambda}; \bm{\theta} \right]   = \frac{1}{2\eta _0 \xi }
  \sum\limits_{n = 1}^{N} \left\{  S_{0}( t_{n})
  + \eta _1 \, \xi ^{2} S_{1}(t_{n})
  + \xi^{4}  S_{2}(t_{n}) \right\},
\end{equation}

\noindent where $S_{0}(t_n)= X^{2}_{\lambda} (t_n )$, $S_{1}(t_n)=
\left[ (X_{\lambda} (t_n + \alpha) - X_{\lambda} (t_n))/\alpha
\right]^{2}$ is the square of the forward-difference gradient
approximation, and $S_{2}(t_n)= \left[ X_{\lambda} (t_n + \alpha) +
X_{\lambda} (t_n-\alpha) -2 X_{\lambda}(t_n) \right]^{2}/\alpha^{4}$
is the square of the discrete approximation of the Laplacian. The 1D spectral
density is given by the following expression

\begin{equation}
\label{covspd1} \tilde{G}_{\rm{x}}(k; \bm{\theta})=
 \frac{ \tilde K_\lambda  (k)  \,\eta _0 \,\xi  }{1 + \eta _1 \,(k \xi )^2  + (k \xi )^4 },
\end{equation}

\noindent where $\tilde K_\lambda  (k)$ is the Fourier transform
(FT) of the smoothing kernel that imposes the resolution $\lambda$.
The autocovariance function is obtained from the inverse FT, i.e,.
by the integral

\begin{equation}
\label{cov1d} G_{\rm x}(t; \bm{\theta})={\int}_{-\infty}^{\infty}
\frac{dk}{2\pi}
              \tilde{G}_{\rm x}(k; \bm{\theta})\exp({\jmath k t}).
\end{equation}

\section{Parameter Inference: The Modified Method of Moments}
\label{mmom}

The modified method of moments (MMoM) is based on fitting sample
constraints with corresponding stochastic ones. The former are based
on the short-range moments $S_{m}(t_{n}),\ m=0,1,2$, appearing in
Eq.~(\ref{H1dS}). They are evaluated by means of sample averages:
$\overline{S_{0}(t_n)}$, $\overline{S_{1}(t_n)}$, and
$\overline{S_{2}(t_n)}$. The respective stochastic constraints can
be expressed as follows:

\begin{equation}
\label{s01d} E[S_0]=\rm{G}_{x}(0)=\frac{\eta_0
\xi}{\pi}{\int}_{0}^{k_c}dk \frac{1}{1 + \eta _1 \,(k \xi )^2 + (k
\xi )^4 },
\end{equation}

\begin{equation}
\label{s11d} E[S_1]=\frac{2}{\alpha^2} \left[ \rm{G}_{x}(0)
-\rm{G}_{x}(\alpha) \right]=\frac{2 \eta_0 \xi}{\pi
\alpha^2}{\int}_{0}^{k_c}dk \frac{[1-\cos(k\alpha)]}{1 + \eta _1
\,(k \xi )^2  + (k \xi )^4 },
\end{equation}

\begin{equation}
\label{s21d} E[S_2]=\frac{2}{\alpha^4} \left[ 3\rm{G}_{x}(0) +
\rm{G}_{x}(2\alpha) -4 \rm{G}_{x}(\alpha)\right]=\frac{2 \eta_0
\xi}{\pi \alpha^4}{\int}_{0}^{k_c}dk
\frac{[3+\cos(2k\alpha)-4\cos(k\alpha)]}{1 + \eta _1 \,(k \xi )^2  +
(k \xi )^4 }.
\end{equation}
The stochastic constrains are related to the SRP model and depend on
$\alpha$, $\eta_0$, $\eta_1$, $\xi$, and $k_c$. For the uniform
sampling step $\alpha$ used herein  an infinite  $k_c $
(unlimited-band), will be considered, allowing the stochastic
constraints Eqs. (\ref{s01d}-\ref{s21d}) to be be expressed in
closed form. The model parameters are estimated by minimizing the
following distance metric \cite{dth03}

\begin{equation}
\label{dm} \Phi_{s}[X(s)]=\Bigg |1-\sqrt{\frac{\overline{S_{1}}}{\overline{S_{0}}}\frac{E[S_{0}]}{E[S_{1}]}}\Bigg|^2+
\Bigg|1-\sqrt{\frac{\overline{S_{2}}}{\overline{S_{1}}}\frac{E[S_{1}]}{E[S_{2}]}}\Bigg|^2.
\end{equation}


\section{FCG model based Spartan predictor}
\label{SP} Let us assume that $T_m=\{t_1,\ldots,t_K\}$ is a set of
sampling times and $X({T_m})=\{X(t_1),\ldots,X(t_K)\}$ the set of
measurements. The spacing of $T_m$ is either uniform (full series)
or non-uniform (training subset). We assume that
$T_p=\{z_1,\ldots,z_k\}$ is a set of estimation points, disjoint
from $T_m$, and $\hat{X}(T_p)$ are the estimated values (temporal
predictions). $T=T_m \cup T_p$ is the full measurement - prediction
set with cardinality $N=K+k$.

\paragraph{Interpolation:} \label{interpolation}

The Spartan predictor (SP) is based on maximizing the conditional
probability density $f_X [X(T_p) |X(T_m) ]$. Considering the
relation $f_X [X(T_p) |X(T_m) ] = f_X [X(T)]/f_X [X(T_m) ]$, the
problem reduces to finding the maximum of $f_X [X(T)]$. We
accomplish this by replacing of $H_{\rm{fgc}}$ with  the estimator
$\label{hatH1} \hat{H}_{{\rm{fgc}}}\left[ X(T) ; \bm{\theta}
\right]=\frac{1}{2}X'(T)J_{{\rm{x}}} (\bm{\theta})X(T), $ and
solving the linear system

\begin{equation}
\label{lin_eqs} {\partial \hat{H}_{\rm{fgc}}[ X(T); \bm{\theta} ]} \big/
{\partial X(z_l)} \Bigg |_{\hat{X}(z_l)} = 0, \ \ \ l=1,\ldots,k.
\end{equation}
$\hat{H}_{\rm{fgc}}[ X(T); \bm{\theta} ]$ involves the sampling points as
well as the prediction points, and $J_{{{\rm{x}}}}({\bm{\theta}})$
only depends on the model parameters, not the data. Neglecting
{\it{interactions}} between the prediction points, the linear
predictor can be expressed {\it{explicitly}} by

\begin{equation}
\label{lin_pred} \hat{X}(T_p) = -{\sum}_{L \in V}{J_{{{\rm{x}}
}}(T_L,T_p)}\,  X(T_L) \big/ {J_{{{\rm{x}}}}(T_L,T_L)},
\end{equation}

\noindent where $V$ is the interaction neighborhood. Herein, the
latter extends up to the second-nearest neighbor. The numerical
complexity of SP involved in solving simultaneously $k$ coupled
Eqs.~(\ref{lin_eqs}) is $O(k^3)$. The predictor given by
Eq.~(\ref{lin_pred}) is explicit.

\paragraph{Extrapolation:} \label{extrapolation} Given the set of
measurements $X({T_m})=\{X(t_1),\ldots,X(t_N)\}$  at times
$T_m=\{t_1,\ldots,t_N\}$, we aim to estimate $k$ future values,
$\hat{X}(T_p)$ $p=N+1,\ldots,N+k$. Two approaches are possible: (i)
Multipoint (MP) extrapolation, i.e. solving simultaneously  the
system of equations~(\ref{lin_eqs}), where $l=N+1,\ldots,N+k,$
%
or (ii) iterative feed-forward (IFF) point-like prediction. The
latter is based on the \textit{short-range memory} property $f_X
[X(t_{l}) |X(\{t_{l-3},t_{l-2},t_{l-1}\})] = f_X [X(t_{l})
|X(\{t_{l-2},t_{l-1}\})]$, which allows the following
{\it{explicit}} predictor, where $\hat{X}(t_{i}) = X(t_{i})$ for
$i=N-1,N:$

\begin{equation}
\label{est} \hat{X}(t_{l}) = -\left[ {J_{l-2,l}}
\hat{X}(t_{l-2})+{J_{l-1,l}} \hat{X}(t_{l-1}) \right] \Big/
{J_{l,l}}\ \ \ l=N+1,\ldots,N+k.
\end{equation}

\section{Description of Data and Methods of Analysis}
\label{simul}

The time series used in this study consists of 388 quarterly
($\alpha$ = 1/4 year) S\&P 500 index data, recorded in 1900-1996
\cite{makri98}. The Spartan parameters are estimated by means of the
MMoM and the maximum likelihood estimation (MLE) method
\cite{fish12}, using a training sets of 132 points. These are
randomly selected from the 388 points to obtain 100 different
configurations. The optimization uses the Nelder-Mead simplex search
algorithm \cite{press} and is terminated when both the model
parameters and the cost function change between consecutive steps
less than $\epsilon = 10^{-6}$. The cost function is given by the
negative log-likelihood -NLL- function in the case of MLE, and the
distance metric -DM-, given by Eq.~(\ref{dm}), in the MMoM case.
Initial guesses for the Spartan model parameters are
$\xi^{(0)}_{i}=\alpha$ and $\eta_1^{(0)}=1$.

To evaluate the prediction performance, for each realization of the
training set, the remaining 256 points (validation set) are
predicted by both SP (MP and IFF) and
Kolmogorov-Wiener predictor (KWP) \cite{cres93}. In both cases, the
Spartan covariance model is used. In KWP, the search neighborhood
includes the entire series.
The following statistics are evaluated, where $X_i$ is the real
value, $\hat{X}_i$ is the estimate, and $M$ is the number of
validation points: (i) mean absolute error (MAE):
$\sum_{i=1}^{M}|X_i - \hat{X_i}|/M$ (ii)  mean relative error (MRE):
$\sum_{i=1}^{M}(X_i - \hat{X_i})/{X_i}$ (iii) mean absolute relative
error (MARE): $\sum_{i=1}^{M} \big|(X_i - \hat{X_i})/M \, X_i \big|$
(iv) root mean square error (RMSE): $\sqrt{\sum_{i=1}^{M}|X_i -
\hat{X_i}|^2/M}$, and (v) the linear correlation coefficient (R).
The computations are performed in the Matlab$\circledR$ environment
on a desktop computer with a Pentium 4 CPU at 3 GHz and 1 GB of RAM.

\section{Results}
\label{res}

\begin{figure}
  \hfill
  \begin{minipage}[t]{.45\textwidth}
  \begin{center}
\includegraphics[scale=0.4]{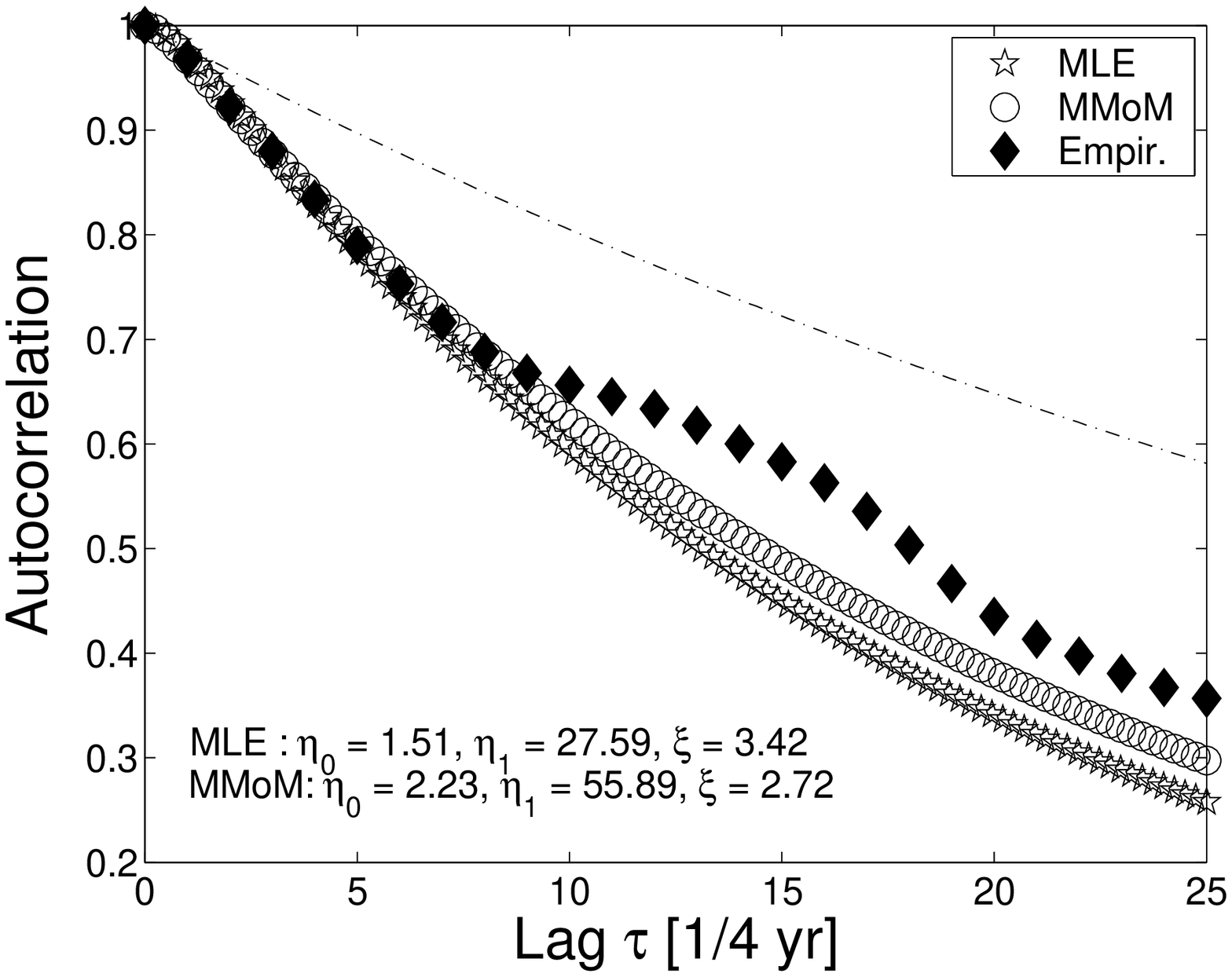}
\caption{\label{fig:1}Empirical correlation function ($\lozenge$)
and optimal Spartan model obtained from complete data by means of
MLE ($\star$) and MMoM ($\circ$); for dash-dot line, see text.}
\end{center}
  \end{minipage}
  \hfill
  \begin{minipage}[t]{.45\textwidth}
  \begin{center}
\includegraphics[scale=0.4]{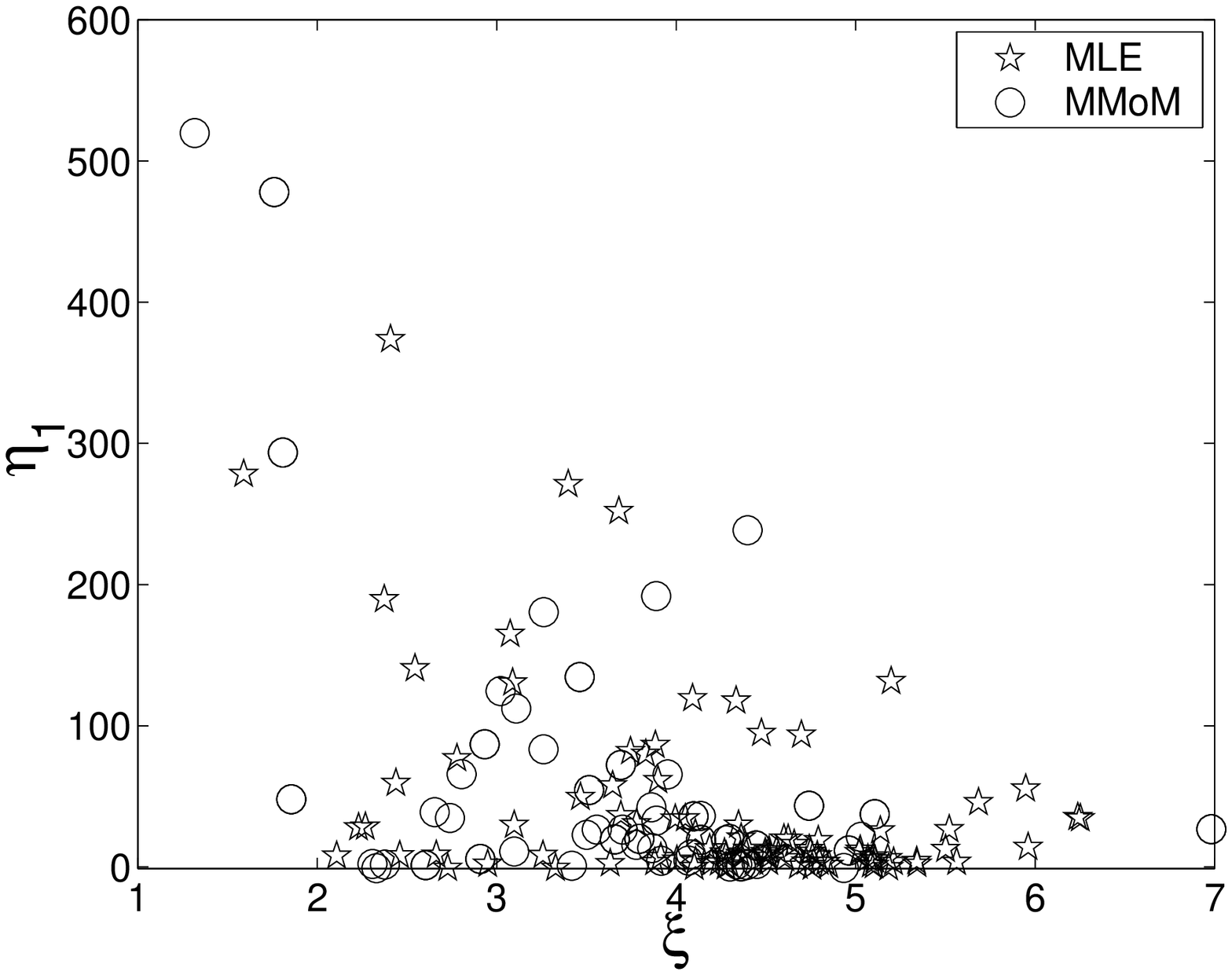}
\caption{\label{fig:2}Estimates of $\xi$ and $\eta_1$, obtained from
100 training sets of 132 points by MMoM ($\circ$) and MLE ($\star$).
Note: in the case of MMoM some estimates lie outside the axes'
range.}
  \end{center}
  \end{minipage}
  \hfill
\end{figure}

\paragraph{Estimation of Correlations:}
The empirical  correlation function is compared with those obtained
with the MMoM and the MLE estimators in Fig.~\ref{fig:1}. They all
match very well near the origin, which is crucial for interpolation.
The value of the MMoM distance metric function is $\Phi=8.2 \times
10^{-20}$, indicating excellent match of the sample and stochastic
constraints. The optimization CPU time of MMoM (0.078 s) is 228
times faster than the MLE one (17.08 s). Unlike the MLE CPU time
that increases nonlinearly with the data, the MMoM CPU time is
insensitive to the domain size \cite{dth03}.  The difference between
the two methods in computational time is expected to increase
dramatically with the sample size.

As shown in Fig. \ref{fig:2}, the distribution of parameter
estimates, obtained from different training sets, is scattered and skewed for both MMoM and MLE cases. Some MMoM estimates receive extreme values, e.g. $\eta_0 > 10^2$, $\eta_1
> 10^3$, and $\xi < 10^{0}$ (not shown in Fig. \ref{fig:2}),
producing an almost  linear decrease of the correlations (e.g.,
dash-dot line in Fig. \ref{fig:1} corresponds to $\xi=0.065,
\eta_1=5 \times 10^{5}$). Such cases are marked by relatively high
 $\Phi$ values, indicating  poor  matching of the sample
and stochastic moments, likely due to non-ergodic conditions.
Indeed, the integral scale \cite{hris-elog06} is quite large
$I(\eta_1=55.89,\xi=2.72)=41.4$, compared to the time series'
length.

\begin{table}
\caption{\label{tab.1}Interpolation errors based on 100 training
sets of 132 points.}
\begin{ruledtabular}
\begin{tabular}{lccccc}
           &        MAE &       MARE &        MRE &       RMSE &       R \\
\hline
       MMoM &     0.0717 &     0.0763 &    -0.0168 &     0.1174 &     0.9519\\
       MLE &     0.0731 &     0.0771 &    -0.0140 &     0.1202 &     0.9493\\
\end{tabular}
\end{ruledtabular}
\end{table}

\begin{figure}
\includegraphics[scale=.4]{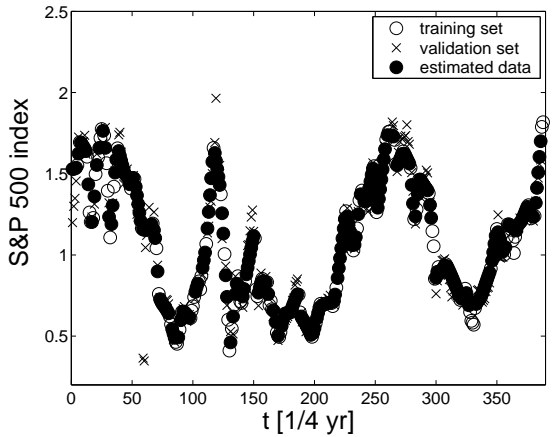}
\caption{\label{fig:3} Time series reconstruction by interpolation:
training ($\circ$), validation ($\times$), estimates ($\bullet$).}
\end{figure}

\paragraph{Interpolation:}
 In the absence of long-range correlations, the short distances are
most relevant  for interpolation. This is evidenced from Table
\ref{tab.1}, which compares the interpolation errors of the MLE- and
MMoM-based estimates: the non-ergodic effect does not introduce
significant deviations in the MMoM-based estimates. We evaluate the
performance of the interpolation methods by comparing the
observations at the validation points with the predictions, using
the statistical measures of performance defined above. The model
parameters estimates are based on the MMoM. In Table \ref{tab:2},
the SP prediction errors are compared with those of KWP. The
comparison distinguishes between areas with different densities of
training data.  Nine categories are defined according to the number
of training-point neighbors in the vicinity of the validation point.
The category $(i,j)$ implies points from the validation set with $i$
nearest and $j$ next-nearest neighbors that belong in the training
set. Naturally, the categories with more data in their interaction
neighborhood display smaller errors than those with fewer data.
There are no significant differences between the SP and KWP results.
The linear correlation coefficient values (0.952 for SP and 0.953
for KWP) indicate strong correlation between the predictions and the
actual data. An example of a reconstructed time series from 132
points, obtained  by means of SP  is shown in Fig. \ref{fig:3}.

\begin{table}[h]
\caption{\label{tab:2}Interpolation errors of SP and KWP predictions
based on 100 training configurations.}
\begin{ruledtabular}
\begin{tabular}{lcccccccccc}
&      (2,2) &      (1,2) &      (0,2) &      (0,1) &      (0,0) &      (1,1) &      (1,0) &      (2,0) &      (2,1) &      Total\\
\hline
           &                                                                    \multicolumn{ 9}{c}{{\bf Mean Absolute Error}} \\
\hline
{\bf SP} &      0.041 &      0.050 &      0.063 &      0.085 &      0.126 &      0.051 &      0.053 &      0.036 &      0.034 &     0.0718 \\

{\bf KWP} &      0.042 &      0.051 &      0.063 &      0.085 &      0.123 &      0.052 &      0.053 &      0.036 &      0.034 &     0.0717\\
\hline
           &                                                           \multicolumn{ 9}{c}{{\bf Mean Absolute Relative Error}} \\
\hline
{\bf SP} &      0.038 &      0.052 &      0.067 &      0.097 &      0.129 &      0.054 &      0.059 &      0.039 &      0.033 &     0.0756\\

{\bf KWP} &      0.040 &      0.053 &      0.067 &      0.097 &      0.126 &      0.055 &      0.059 &      0.039 &      0.033 &     0.0761\\
\hline
           &                                                                    \multicolumn{ 9}{c}{{\bf Mean Relative Error}} \\
\hline
{\bf SP} &     -0.001 &     -0.005 &     -0.010 &     -0.034 &     -0.032 &     -0.010 &     -0.016 &     -0.003 &     -0.001 &    -0.0168\\

{\bf KWP} &     -0.001 &     -0.004 &     -0.009 &     -0.034 &     -0.032 &     -0.010 &     -0.015 &     -0.002 &      0.000 &    -0.0160\\
\hline
           &                                                                 \multicolumn{ 9}{c}{{\bf Root Mean Square Error}} \\
\hline
{\bf SP} &      0.051 &      0.073 &      0.089 &      0.130 &      0.174 &      0.082 &      0.084 &      0.052 &      0.049 &     0.1174\\

{\bf KWP} &      0.053 &      0.075 &      0.089 &      0.130 &      0.165 &      0.082 &      0.084 &      0.053 &      0.050 &     0.1162\\
\end{tabular}
\end{ruledtabular}
\end{table}

\paragraph{Extrapolation:}
In Table \ref{tab:3} we compare SP errors, obtained by the SP-MP and
SP-IFF methods with those obtained using a second-order
autoregressive (AR) model. We randomly select from the $N=388$ data
a point $i$ ($2<i<N-k$) and predict the following $k=3$ values. The
errors obtained by the two SP methods are almost identical and
comparable  (slightly smaller) with those obtained from the AR
model. At 1-st, 2-nd and 3-rd lag, the predicted future values have
average relative errors on the order of 5\%, 10\%, and 15\%,
respectively.

\begin{table}
\caption{\label{tab:3}Extrapolation error statistics of SP (MP and
IFF) and AR model based on 100 randomly selected triplets
$i+1,i+2,i+3$ ($2<i<N-3$).}
\begin{ruledtabular}
\begin{tabular}{lccccccccc}
{{\bf Method}}           &   \multicolumn{ 3}{c}{{\bf SP-MP}} & \multicolumn{3}{c}{{\bf SP-IFF}} & \multicolumn{3}{c}{{\bf AR}}\\
\hline
{\bf{Lag $\tau$}} &          1 &          2 &          3 &          1 &          2 &          3 &          1 &          2 &          3 \\
\hline
       {\bf{MAE}} &    0.0549   &    0.0937 &    0.1390 &    0.0548 &    0.0937 &   0.1390 &    0.0583 &    0.1068 &    0.1549\\

      {\bf{MARE}} &    0.0495 &    0.0833 &    0.1383 &    0.0495 &    0.0832 &    0.1384 &    0.0533 &    0.0996 &    0.1633\\

      {\bf{MRE}} &    0.0047 &    0.0074 &    -0.0156 &    0.0048 &    0.0075 &    -0.0155 &    -0.0096 &    -0.0211 &    -0.0583\\

     {\bf{RMSE}} &    0.0696 &    0.1164 &    0.1857 &    0.0695 &    0.1164 &   0.1858 &    0.0827 &    0.1396 &    0.2110\\
\end{tabular}
\end{ruledtabular}
\end{table}

\section{Summary}
\label{sum}

We present a framework for the analysis of time series  based on
`pseudo-energy' functionals that  capture the temporal heterogeneity
of the observed process. Estimates of the process at unmeasured
points (predictions) are based on the mode of the joint pdf. The
Spartan prediction method is tested on financial data (S\&P 500
index time series), by both interpolation and extrapolation. The SP
yields results well comparable to those obtained by the standard
Kolmogorov-Wiener predictor at generally lower computational cost.

We also present an efficient parameter inference technique -
modified method of moments (MMoM), which is based on fitting of
sample and corresponding stochastic short-range constraints. The
main advantage of MMoM is low computational complexity (high speed)
and independence on the domain size, which makes it suitable for
large data sets, difficult to manage by other techniques.
Alternatively, the computational ease gives the method potential
include time-dependent parameters that are continuously estimated,
in a "moving window" approach, to account for potential
non-stationarity. Since the Gaussian assumption is often
unjustified, we currently focus on formulating  a Spartan model
capable of representing directly non-Gaussian data, without the need
for a normalizing transformation.

\begin{acknowledgments}
This work is partially supported by the Marie Curie Transfer of Knowledge Program,
Project SPATSTAT 014135.
\end{acknowledgments}


\end{document}